\documentclass{basi}
\usepackage{txfonts}
\usepackage{graphicx}

\begin{document}
\title[Black hole physics today]{Key problems in black hole physics today}
\author[Pankaj S. Joshi]%
       {Pankaj S. Joshi,\thanks{e-mail: psj@tifr.res.in}\\
        Tata Institute of Fundamental Research, Homi Bhabha Road, Mumbai 400005, India}

\pubyear{2011}
\volume{39}
\pagerange{\pageref{firstpage}--\pageref{lastpage}}
\date{Received 2011 April 04; accepted April 14}
\maketitle
\label{firstpage}

\begin{abstract}
We review here some of the major open issues 
and challenges in black hole physics today, and the 
current progress on the same. 
It is pointed out that to secure a concrete foundation 
for the basic theory as well as astrophysical applications 
for black hole physics, it is essential to gain a 
suitable insight into these questions. In particular, 
we discuss the recent results investigating the final 
fate of a massive star within the framework of the 
Einstein gravity, and the stability and genericity aspects 
of the gravitational collapse outcomes in terms of 
black holes and naked singularities. 
Recent developments such as spinning up a black hole 
by throwing matter into it, and physical effects 
near naked singularities are considered. It is pointed 
out that some of the new results obtained in recent 
years in the theory of gravitational collapse imply 
interesting possibilities and understanding for the 
theoretical advances in gravity as well as towards
new astrophysical applications.  

%
%\verb+http://www.ncra.tifr.res.in/~basi/+.
%
\end{abstract}

%------------------------------------------------------------------------------%
% main text of the paper, using \section, \subsection, \subsubsection          %
%------------------------------------------------------------------------------%

\section{Introduction}\label{s:intro}

The fundamental question of the final fate of a massive star, 
when it exhausts its internal nuclear fuel and collapses continually
under the force of its own gravity, was highlighted by Chandrasekhar 
way back in 1934 (Chandrasekhar 1934), who pointed out: 

``Finally, it is necessary to emphasize one
major result of the whole investigation, namely, that the 
life-history of a star of small mass must be essentially different from
the life-history of a star of large mass. For a star of small mass
the natural white-dwarf stage is an initial step towards complete
extinction. A star of large mass ($ > M_c$) cannot pass
into the white-dwarf stage, and one is left speculating on
other possibilities.'' 

We can see the seeds of modern black hole physics 
already present in the inquiry made above on the final fate 
of massive stars. The issue of endstate of large mass stars has, 
however, remained unresolved and elusive for a long time 
of many decades after that. In fact, a review of the status
of the subject many decades later notes, ``Any stellar core
with a mass exceeding the upper limit that undergoes
gravitational collapse must collapse to indefinitely high
central density... to form a (spacetime) singularity'' 
(Report of the Physics Survey Committee 1986).

The reference above is to the prediction by general 
relativity, that under reasonable physical conditions,
the gravitationally collapsing massive star must terminate 
in a spacetime singularity (Hawking \& Ellis 1973). 
The densities, spacetime curvatures, 
and all physical quantities must typically go to 
arbitrarily large values close to such a singularity. 
The above theoretical result on the 
existence of singularities is, however, of a rather general
nature, and provides no information on the nature and
structure of such singularities. In particular, it 
gives us no information as to whether such singularities,
when they form, will be covered in horizons of gravity
and hidden from us, or alternatively these could be visible to external 
observers in the Universe.

One of the key questions in black hole physics 
today therefore is, are such singularities resulting from collapse,
which are super-ultra-dense regions forming in spacetime, 
visible to external observers in the Universe? This is 
one of the most important unresolved issues in gravitation 
theory currently. Theorists generally believed that in such 
circumstances, a black hole will always form covering the 
singularity, which will then be always hidden from external 
observers. Such a black hole is a region of spacetime from 
which no light or particles can escape. The assumption 
that spacetime singularities resulting from collapse would be 
always covered by black holes is called the 
Cosmic Censorship Conjecture (CCC; Penrose 1969). 
As of today, we do not have any proof or any specific 
mathematical formulation of the CCC available within the
framework of gravitation theory.

If the singularities were always covered in horizons and if 
CCC were true, that would provide a much needed basis for the 
theory and astrophysical applications of black holes. On the 
other hand, if the spacetime singularities which result from a 
continual collapse of a massive star were visible to external
observers in the Universe, we would then have the opportunity to
observe and investigate the super-ultra-dense regions in the 
Universe, which form due to gravitational collapse and where 
extreme high energy physics and also quantum gravity 
effects will be at work.

My  purpose here is to review the above and some of 
the related key issues in gravitation theory and black hole 
physics today. This will be of course from a perspective 
of what I think are the important problems, and no claim 
to completeness is made. In Section 2, we point out 
that in view of the lack of any theoretical progress on CCC, 
the important way to make any progress on this problem 
is to make a detailed and extensive study of gravitational 
collapse in general relativity. Some recent progress
in this direction is summarized. While we now seem to have 
a good understanding of the black hole and naked singularity 
formations as final fate of collapse in many 
gravitational collapse models, 
the key point now is to understand the genericity and stability 
of these outcomes, as viewed in a suitable framework. 
Section 3 discusses these issues in some detail. Recent 
developments on throwing matter into a black hole and the effect 
it may have on its horizon are pointed out in Section 4,
and certain quantum aspects are also discussed. The issue 
of predictability or its breakdown in gravitational collapse 
is discussed in Section 5. We conclude by giving a brief 
idea of the future outlook and possibilities in the 
final section.

\section{What is the final fate of a massive star?}\label{s:final fate}

While Chandra's work pointed out the stable configuration limit
for the formation of a white dwarf, the issue of the final fate of a star
which is much more massive (e.g. tens of solar masses) remains
very much open even today. Such a star cannot settle either as
a white dwarf or as a neutron star.

The issue is clearly important both in high energy astrophysics
and in cosmology. For example, our observations today on the 
existence of dark energy in the Universe and its acceleration 
are intimately connected to the the observations of Type Ia supernovae
in the Universe. The observational evidence coming from
these supernovae, which are
exploding stars in the faraway Universe, tells us 
on how the Universe may be accelerating away and the rate 
at which such an acceleration is taking place. While Type Ia
supernovae result from the explosion of a white dwarf star, 
at the heart 
of a Type II supernova underlies the phenomenon of a 
catastrophic gravitational collapse of a massive star, 
wherein a powerful shock wave is generated, blowing 
off the outer layers of the star.

If such a star is able to throw away enough of matter in
such an explosion, it might eventually settle as a neutron star.
But otherwise, or if further matter is accreted onto
the neutron star, there will be a further continual collapse,
and we shall have to then explore and investigate 
the question of the final fate of such a massive 
collapsing star. But other stars, which are more
massive and well above the normal supernova mass limits
must straightaway enter a continual collapse mode
at the end of their life cycle,
without an intermediate neutron star stage. The final
fate of the star in this case must be decided by
general relativity alone.  

The point here is, more massive stars which are tens of 
times the mass of the Sun burn much faster and are far more 
luminous. Such stars then cannot survive more than about ten
to twenty million years, which is a much shorter life span 
compared to stars like the Sun, which live billions of years.
Therefore, the question of the final fate of such short-lived 
massive stars is of central importance in astronomy
and astrophysics.

What happens then, in terms of the final outcome, when
such a massive star dies after exhausting its internal nuclear fuel? 
As we indicated above, the general theory of relativity
predicts that the collapsing massive star must 
terminate in a spacetime singularity, where the matter
energy densities, spacetime curvatures and other physical 
quantities blow up.
It then becomes crucial to know whether such super-ultra-dense
regions, forming in stellar collapse, 
are visible to an external
observer in the Universe, or whether they will be 
always hidden within a black hole and an event horizon 
that could form as the star collapses.    
This is one of the most important issues in black hole
physics today.

The issue has to be probed necessarily within the framework of
a suitable theory of gravity, because the strong gravity effects 
will be necessarily important in such a scenario. This was
done for the first time in the late 1930s, by the works of
Oppenheimer and Snyder, and Datt 
(Oppenheimer \& Snyder 1939; Datt 1938).
They used the general theory 
of relativity to examine the final
fate of an idealized massive matter cloud, which was taken to be 
a spatially homogeneous ball which had no rotation or internal
pressure, and was assumed to be spherically symmetric.
The dynamical collapse studied here resulted in the formation
of a spacetime singularity, which was preceded by the development
of an event horizon, which created a black hole in the spacetime.
The singularity was hidden inside such a black hole, and the 
collapse eventually settled into a final state which was
the Schwarzschild geometry (see Fig. 1).

\begin{figure}
\centerline{\includegraphics[width=9cm]{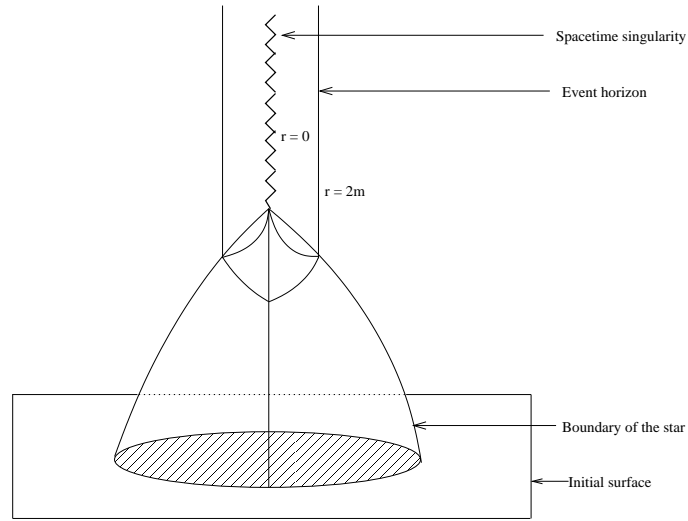}}
\caption{Dynamical evolution of a homogeneous
spherical dust cloud collapse, as described by
the Oppenheimer-Snyder-Datt solution. \label{f:one}}
\end{figure}

There was, however, not much attention paid to this model
at that time, and it was widely thought by gravitation theorists 
as well as astronomers that it would be absurd for a star 
to reach such a final ultra-dense state of its evolution.  
It was in fact only as late as 1960s, that a resurgence of 
interest took place in the topic, when important 
observational developments in astronomy
and astrophysics revealed several very high energy 
phenomena in the Universe, such as quasars and radio galaxies,
where no known physics was able to explain
the observations of such extremely high energy phenomena
in the cosmos. Attention was drawn then to dynamical 
gravitational collapse and its final fate, and in fact 
the term `black hole' was also popularized just around the 
same time in 1969, by John Wheeler.

The CCC also came into existence in 1969. It suggested
and assumed that what happens in the Oppenheimer-Snyder-Datt (OSD) 
picture of gravitational collapse, as discussed above, would be 
the generic final fate of a realistic collapsing massive 
star in general. In other words, it was assumed that 
the collapse of a realistic massive star will terminate
in a black hole, which hides the singularity,
and thus no visible or naked singularities will develop
in gravitational collapse. 
Many important developments then took 
place in black hole physics
which started in earnest, and several important theoretical 
aspects as well as astrophysical applications of black holes
started developing. The classical as well as quantum aspects
of black holes were then explored and interesting 
thermodynamic analogies for black holes were also developed.  
Many astrophysical applications for the real Universe
then started developing for black holes, such as making 
models using black holes for phenomena such as jets from 
the centres of galaxies and the extremely energetic gamma 
rays bursts.

The key issue raised by the CCC, however, still 
remained very much open, namely whether a real star will
necessarily go the OSD way for its final state of collapse,
and whether the final singularity will be always necessarily
covered within an event horizon. This is     
because real stars are inhomogeneous, have internal pressure
forces and so on, as opposed to the idealized OSD assumptions.
This remains an unanswered question, which is one
of the most important issues in gravitation physics and 
black hole physics today.
A spacetime singularity that is visible to faraway
observers in the Universe is called a naked singularity
(see Fig. 2).
The point here is, while general relativity predicts the 
existence of singularity as the endstate for collapse, it gives
no information at all on the nature or structure of such
singularities, and whether they will be covered by 
event horizons, or would be visible to external 
observers in the Universe. 

\begin{figure}
\centerline{\includegraphics[width=9cm]{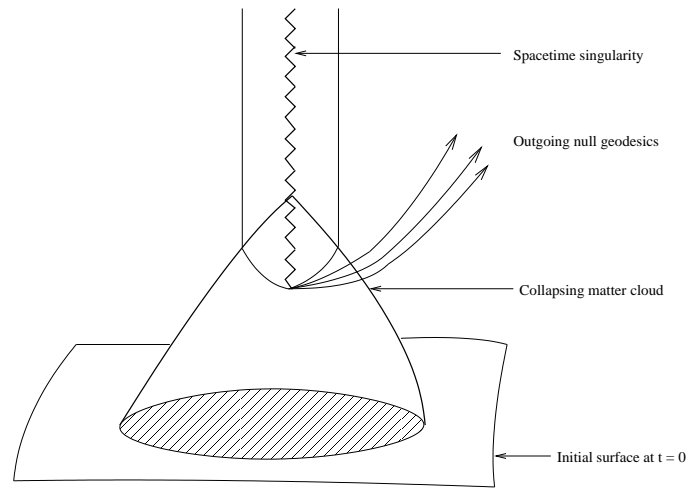}}
\caption{A spacetime singularity of gravitational collapse
which is visible to external observers in the Universe,
in violation to the cosmic censorship conjecture. \label{f:one}}
\end{figure}

There is no proof, or even any mathematically 
rigorous statement available for CCC after many decades 
of serious effort. What is really needed to resolve the 
issue is gravitational collapse models for a 
realistic collapse configuration, with inhomogeneities 
and pressures included. The effects need to be worked out and 
studied in detail within 
the framework of Einstein gravity. Only such considerations
will allow us to determine 
the final fate of collapse in terms of either 
a black hole or a naked singularity final state.

Over the past couple of decades, many such 
collapse models have been worked out and studied in
detail. The generic conclusion that has emerged 
out of these studies is that both the black holes 
and naked singularity final states do develop 
as collapse endstates, for a realistic gravitational 
collapse that incorporates inhomogeneities as well as 
non-zero pressures within the interior of the collapsing
matter cloud. Subject to various regularity and
energy conditions to ensure the physical reasonableness
of the model, it is the initial data, in terms of the 
initial density, pressures, and velocity profiles for
the collapsing shells, that determine the final fate
of collapse as either a naked singularity or a 
black hole (for further detail and references 
see e.g. Joshi 2008).

\section{The genericity and stability of collapse outcomes}
While general relativity may predict the 
existence of both black holes and naked singularities 
as collapse outcome, an important question then is, 
how would a realistic continual gravitational collapse 
of a massive star in nature would end up. 
Thus the key issue under active debate now 
is the following: 
Even if naked singularities did develop as collapse 
end states, would they be generic or stable in some 
suitably well-defined sense, as permitted by the gravitation 
theory? The point here is, if naked singularity formation 
in collapse was necessarily `non-generic' in some 
appropriately well-defined sense, 
then for all practical purposes, a realistic physical 
collapse in nature might always end up in a black hole, 
whenever a massive star ended its life.

In fact, such a genericity requirement has been 
always discussed and desired for any possible mathematical 
formulation for CCC right from its inception. However, 
the main difficulty here has again been that, there 
is no well-defined or precise notion of genericity 
available in gravitation theory and the general theory 
of relativity. Again, it is only various gravitational 
collapse studies that can provide us with 
more insight into this genericity aspect also.

A result that is relevant here is the following
(Joshi \& Dwivedi, 1999; Goswami \& Joshi, 2007).
For a spherical gravitational collapse of a 
rather general (type I) matter field, satisfying the 
energy and regularity conditions, given any regular 
density and pressure profiles at the initial epoch, 
there always exist classes of velocity profiles for the 
collapsing shells and dynamical evolutions as determined 
by the Einstein equations, that, depending on the 
choice made, take the collapse to either a black hole 
or naked singularity final state
(see e.g. Fig. 3 for a schematic illustration of such a scenario).

\begin{figure}
\centerline{\includegraphics[width=9cm]{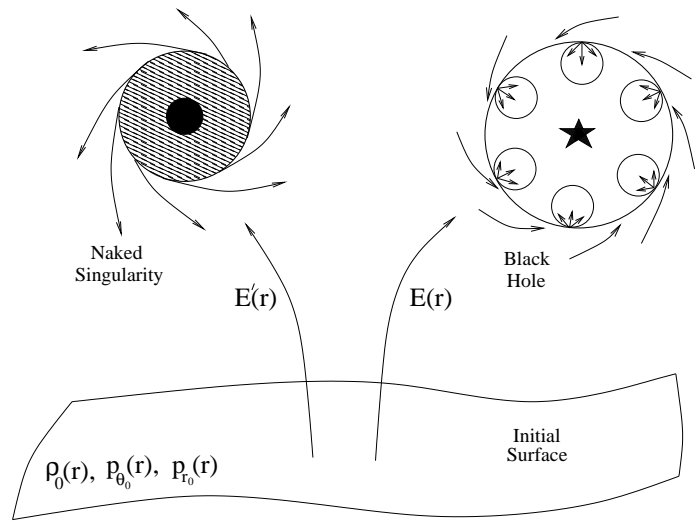}}
\caption{Evolution of spherical collapse for a 
general matter field with inhomogeneities and 
non-zero pressures included. \label{f:three}}
\end{figure}

Such a distribution of final states of collapse
in terms of the black holes and naked singularities 
can be seen much more transparently when we consider
a general inhomogeneous dust collapse,
for example, as discussed by  Mena, Tavakol \& Joshi (2000)
(see Fig.4).

\begin{figure}
\centerline{\includegraphics[width=9cm]{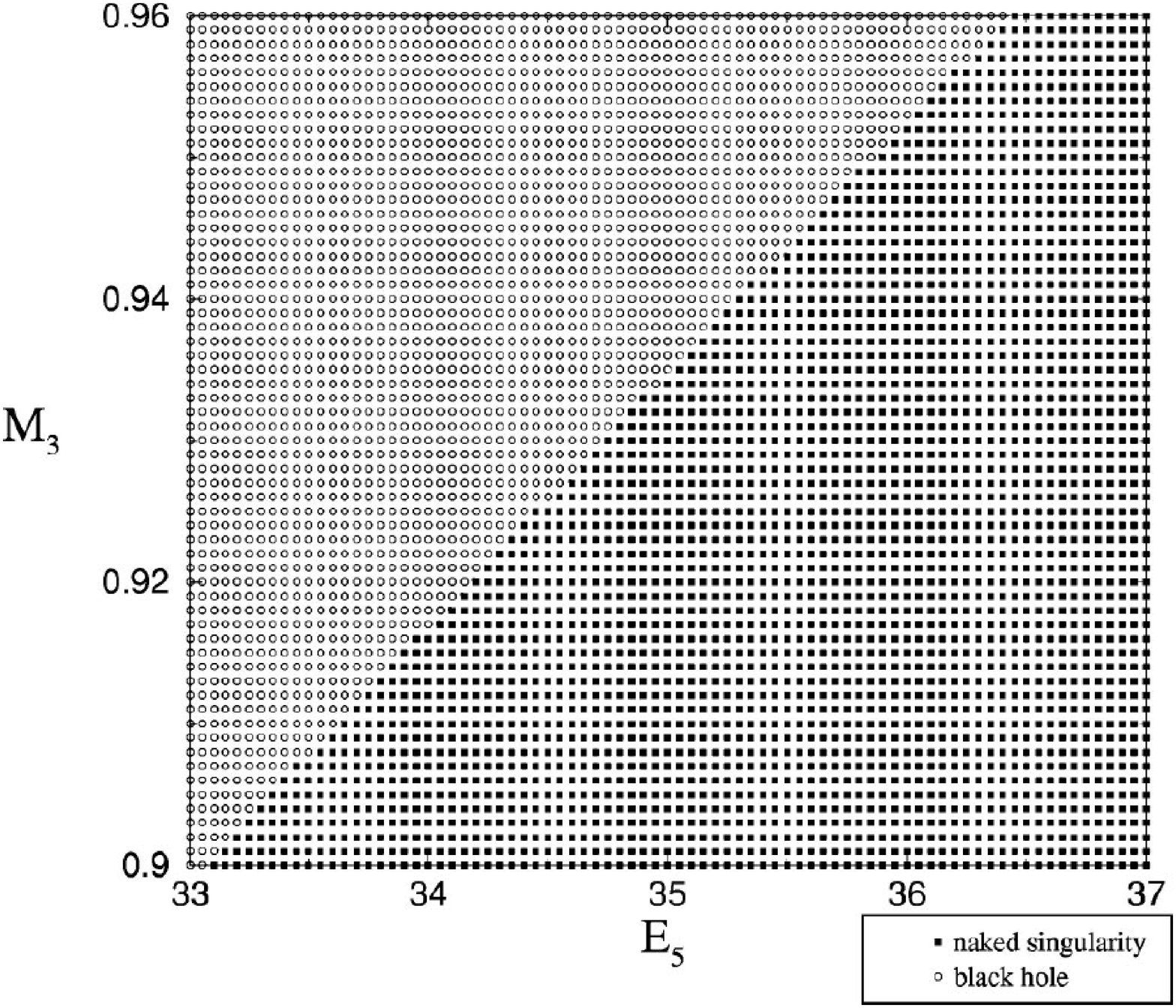}}
\caption{Collapse final states for inhomogeneous
dust in terms of initial mass and velocity profiles
for the collapsing shells. \label{f:four}}
\end{figure}

What determines fully the final fate of collapse 
here are the initial density and velocity profiles
for the collapsing shells of matter.
One can see here clearly how the different choices 
of these profiles for the collapsing 
cloud distinguish between the two 
final states of collapse, and how each of the
black hole and naked singularity states
appears to be `generic' in terms of their being
distributed in the space of final states.
Typically, the result we have here is, given any 
regular initial density profile for the collapsing dust 
cloud, there are velocity profiles that take the collapse
to a black hole final state, and there are other 
velocity profiles that take it to naked singularity
final state. In other words, the overall available 
velocity profiles are divided into two distinct  
classes, namely the ones which take the given density 
profile into black holes, and the other ones that take
the collapse evolution to a naked singularity 
final state. The same holds conversely also, namely 
if we choose a specific velocity profile, then the 
overall density profile space is divided into two 
segments, one taking the collapse to black hole final 
states and the other taking it to naked singularity
final states. The clarity of results here gives us
much understanding on the final fate of a collapsing
matter cloud.

Typically, all stars have a higher density at 
the centre, which slowly decreases as one moves away. 
So it is very useful to incorporate inhomogeneity into 
dynamical collapse considerations. However, much more 
interesting is the collapse with non-zero pressures
which are very important physical forces within a 
collapsing star. We briefly consider below a typical 
scenario of collapse with a non-zero pressure component, 
and for further details we refer to 
Joshi \& Malafarina (2011).

For a possible insight into genericity of naked 
singularity formation in collapse, we investigated 
the effect of introducing small tangential pressure 
perturbations in the collapse dynamics of the classic 
Oppenheimer-Snyder-Datt  gravitational collapse, which is
an idealized model assuming zero pressure, and which 
terminates in a black hole final fate as discussed above.
Thus we study the stability of the OSD black hole under 
introduction of small tangential stresses.

It is seen explicitly that there exist
classes of stress perturbations such that the 
introduction of a smallest tangential pressure within 
the collapsing OSD cloud changes the endstate of collapse 
to formation of a naked singularity, rather than a black hole. 
What follows is that small stress perturbations
within the collapsing cloud change the final fate of
the collapse from being a black hole to a naked singularity. 
This can also be viewed as perturbing the spacetime metric
of the cloud in a small way. Thus we can understand here
the role played by tangential pressures in a well-known 
gravitational collapse scenario. A specific and physically 
reasonable but generic enough class of perturbations is 
considered so as to provide a good insight into the 
genericity of naked singularity formation in collapse
when the OSD collapse model is perturbed by introduction 
of a small pressure. Thus we have an important insight
into the structure of the censorship principle which 
as yet remains to be properly understood.

The general spherically symmetric metric
describing the collapsing matter cloud can be written
as,
\begin{equation}\label{metric}
ds^2=-e^{2\sigma(t, r)}dt^2+e^{2\psi(t, r)}dr^2+R(t, r)^2d\Omega^2,
\end{equation}
with the stress-energy tensor for a generic matter source 
being given by,
$T_t^t=-\rho, \; T_r^r=p_r, \; T_\theta^\theta=T_\phi^\phi=p_\theta$.
The above is a general scenario, in that it involves no assumptions
about the form of the matter or the equation of state.

As a step towards deciding the stability or otherwise 
of the OSD collapse model under the injection of small 
tangential stress perturbations, we consider the dynamical 
development of the collapsing cloud, as governed by the 
Einstein equations. The visibility or otherwise of the 
final singularity that develops in collapse is determined
by the behaviour of the apparent horizon in the spacetime,
which is the boundary of the trapped surface region
that develops as the collapse progresses. First, we define a
scaling function $v(r,t)$ by the relation $R=rv$.
The Einstein equations for the above spacetime geometry
can then be written as,
\begin{eqnarray}\label{p2}
p_r&=&-\frac{\dot{F}}{R^2\dot{R}}, \; \rho = \frac{F'}{R^2R'} \; ,\\ \label{sigma2}
\sigma'&=&2\frac{p_\theta-p_r}{\rho+p_r}\frac{R'}{R}-\frac{p_r'}
{\rho+p_r} \; ,\\ \label{G2}
2\dot{R}'&=&R'\frac{\dot{G}}{G}+\dot{R}\frac{H'}{H} \; ,\\
\label{F2}
F&=&R(1-G+H) \; ,
\end{eqnarray}
The functions $H$ and $G$ in the above 
are defined as,
$H =e^{-2\sigma(r, v)}\dot{R}^2 , \; G=e^{-2\psi(r, v)}R'^2$.
The above are five equations in
seven unknowns, namely $\rho,\; p_r, \; p_{\theta}, \; R,\; F,\; G,\; H$.
Here $\rho$ is the mass-energy density, $p_r$ and 
$p_\theta$ are the radial and tangential stresses respectively, 
$R$ is the physical radius for the matter cloud, and ${F}$ 
is the Misner-Sharp mass function.

It is possible now, with the above definitions 
of $v, H$ and $G$,  to substitute the unknowns $R, H$ 
with $v$, $\sigma$. Then, without loss of generality, the 
scaling function $v$ can be written as $v(t_i, r)=1$ 
at the initial time $t_i=0$, when the collapse begins.
It then goes to zero at the spacetime singularity $t_s$, which
corresponds to $R=0$, and thus we have $v(t_s, r)=0$.
This amounts to the scaling $R=r$ at the initial
epoch of the collapse, which is an allowed freedom. 
The collapse condition here is $\dot R<0$ throughout 
the evolution, and this is equivalent to $\dot{v}<0$.

One can integrate the Einstein equations, at least
up to one order, to reduce them to a first order system,
to obtain the function $v(t,r)$. This function, which is 
monotonically decreasing in $t$ can be inverted to obtain 
the time needed by a matter shell at any radial value 
$r$ to reach the event with a particular value $v$. 
We can then write the function $t(r, v)$ as,
\begin{equation}
t(r, v)= \int^1_{v}\frac{e^{-\sigma}}{\sqrt{\frac{F}{r^3\tilde{v}}
+\frac{be^{2rA}-1}{r^2}}}d\tilde{v} \; .
\end{equation}
The function $A(r,v)$ in the above depends on the 
nature of the tangential stress perturbations chosen.
The time taken by the shell at $r$ to reach the spacetime 
singularity at $v=0$ is then $t_s(r)=t(r, 0)$.

Since $t(r, v)$ is in general at least $C^2$ everywhere in the spacetime
(because of the regularity of the functions involved), and is continuous at the
centre, we can write it as,
\begin{equation}\label{t}
t(r, v)= t(0, v)+r\chi(v)+O(r^2) \;
\end{equation}
Then, by continuity, the time for a shell located at any $r$ close to
the centre to reach the singularity is given as,
\begin{equation}
t_s(r)= t_s(0)+r\chi(0)+O(r^2)
\end{equation}
Basically, this means that the singularity curve should have a well-defined
tangent at the center. Regularity at the center also implies that
the metric function $\sigma$ cannot have constant or linear terms in
$r$ in a close neighborhood of $r=0$, and it must go as  $\sigma\sim r^2$
near the center. Therefore the most general choice of the free function
$\sigma$ is,
\begin{equation}
\sigma(r,v)=r^2g(r,v) \;
\end{equation}
Since $g(r, v)$ is a regular function (at least $C^2$), it can be
written near $r=0$ as,
\begin{equation}\label{expand-g}
g(r, v)=g_0(v)+g_1(v)r+g_2(v)r^2+...
\end{equation}

We can now investigate how the OSD gravitational
collapse scenario, which gives rise to a black hole as
the final state, gets altered when small stress perturbations 
are introduced in the dynamical evolution of collapse.
For that we first note that the dust model is obtained if
$p_r=p_{\theta}=0$ in the above. In that case, 
$\sigma'=0$ and together with the condition $\sigma(0)=0$ gives
$\sigma=0$ identically.
In the OSD homogeneous collapse to a black hole, 
the trapped surfaces
and the apparent horizon develop much earlier before
the formation of the final singularity. 
But when density inhomogeneities are allowed in the 
initial density profile, such as a higher density at the 
centre of the star, then the trapped surface formation is 
delayed in a natural manner within the collapsing cloud.
Then the final singularity becomes visible to 
faraway observers in the Universe
(e.g. Joshi, Dadhich \& Maartens 2002).

The OSD case is obtained from the inhomogeneous dust
case, when we assume further that the collapsing dust 
is necessarily homogeneous at all epochs of collapse. This 
is of course an idealized scenario because realistic
stars would have typically higher densities at the centre, and 
they also would have non-zero internal stresses.
The conditions that must be imposed to obtain the OSD case
from the above are given by $M=M_0\; v=v(t)\; b_0(r)=k$.
Then we have $F'=3M_0r^2$, $R'=v$, the energy density 
is homogeneous throughout the collapse, and the density
is given by $\rho=\rho (t)= {3M_0}/{v^3}$.
The spacetime geometry then becomes the Oppenheimer-Snyder 
metric, which is given by,
\begin{equation}
ds^2=-dt^2+\frac{v^2}{1+kr^2}dr^2+r^2v^2d\Omega^2, \; 
\end{equation}
where the function $v(t)$ is a solution of the equation 
of motion, $ \frac{dv}{dt}=\sqrt{(M_0/v)+k}$,
obtained from the Einstein equation. 
In this case we get $\chi(0)=0$ identically. All the 
matter shells then collapse into a simultaneous singularity, 
which is necessarily covered by the event horizon that 
developed in the spacetime at an earlier time.
Therefore the final fate of collapse is 
a black hole.

To explore the effect of introducing small pressure 
perturbations in the above OSD scenario and to study the 
models thus obtained which are close to the Oppenheimer-Snyder, 
we can relax one or more of the above conditions.
If the collapse outcome is not a black hole, the final
collapse to singularity cannot be simultaneous.  
We can thus relax the condition $v=v(t)$ above, 
allowing for $v = v(t,r)$. We keep the other conditions
of the OSD model unchanged, so as not to depart too much 
from the OSD model, and this should bring out more 
clearly the role played by the stress perturbations 
in the model.
We know that the metric function $\sigma(t,r)$ must 
identically vanish for the dust case. On the other hand, 
the above amounts to allowing for small perturbations in $\sigma$,
and allowing it to be non-zero now. This is equivalent 
to introducing small stress perturbations in the model, 
and it is seen that this affects the apparent horizon 
developing in the collapsing cloud.
We note that taking $M=M_0$ leads to $ F=r^3M_0$.

In this case, in the small $r$ limit we 
obtain $G(r,t)=b(r)e^{2\sigma(r, v)}$.
The radial stress $p_r$ vanishes here as $\dot F=0$, 
and the tangential pressure turns out to have
the form, $p_\theta= p_1 r^2 + p_2 r^3 +...$, where 
$p_1, p_2$ are evaluated in terms of coefficients of 
$m, g$, and $R$ and its derivatives, and we get,
\begin{equation}\label{pt}
p_\theta=3\frac{M_0g_0}{vR'^2}r^2+\frac{9}{2}\frac{M_0g_1}{vR'^2}r^3+...
\end{equation}
As seen above, the choice of the sign of 
the functions $g_0$ and $g_1$ is enough to ensure 
positivity or negativity of the pressure $p_\theta$.
The first order coefficient $\chi$ in
the equation of the time curve of the singularity 
$t_s(r)$ is now obtained as,
\begin{equation}\label{chi}
\chi(0)=-\int^1_0\frac{v^{\frac{3}{2}}g_1(v)}{(M_0+vk+2vg_0(v))^{\frac{3}{2}}}dv \; .
\end{equation}
As we have noted above, it is the quantity 
$\chi(0)$ that governs the nature of the singularity curve, 
and whether it is increasing or decreasing away from 
the center. It can be seen from above that it is the matter 
initial data in terms of the density and stress profiles, 
the velocity of the collapsing shells, and the allowed 
dynamical evolutions that govern and fix
the value of $\chi(0)$.

The apparent horizon in the spacetime and the 
trapped surface formation as the collapse evolves is 
also governed by the quantity $\chi(0)$, which in turn 
governs the nakedness or otherwise of the singularity. 
The equation for the apparent horizon is given by ${F}/{R}=1$. 
This is analogous to that of the dust case
since ${F}/{R}={rM}/{v}$ in both these cases.
So the apparent horizon curve $r_{ah}(t)$ is given by
\begin{equation}\label{ah}
r_{ah}^2=\frac{v_{ah}}{M_0},
\end{equation}
with $v_{ah}=v(r_{ah}(t), t)$, which can also be inverted as a
time curve for the apparent horizon $t_{ah}(r)$.
The visibility of the singularity at the center of 
the collapsing cloud to faraway observers is determined by the
nature of this apparent horizon curve which is given by,
\begin{equation}\label{t-ah}
t_{ah}(r)=t_s(r)-\int_0^{v_{ah}}\frac{e^{-\sigma}}
{\sqrt{\frac{M_0}{v}+\frac{be^{2\sigma}-1}{r^2}}}dv
\end{equation}
where the $t_s(r)$ is the singularity time curve, and its 
initial point is $t_0=t_s(0)$. Near $r=0$ 
we then get,
\begin{equation}
    t_{ah}(r) =t_0+\chi(0)r+o(r^2) \; .
\end{equation}

From these considerations, it is possible to see 
how the stress perturbations affect the time of formation 
of the apparent horizon, and therefore the formation of 
a black hole or naked singularity. A naked singularity 
would typically occur as a collapse endstate when a
comoving observer at a fixed $r$ value does not encounter 
any trapped surfaces before the time of singularity formation. 
For a black hole to form, trapped surfaces must develop 
before the singularity. Therefore it is required 
that,
\begin{equation}
t_{ah}(r) \le t_0 ~~\mbox{for}~~ r>0, ~~\mbox{near}~~ r=0 \; .
\end{equation}
As can be seen from above, for all functions $g_1(v)$ 
for which $\chi(0)$ is positive, this
condition is violated and in that case the apparent 
horizon is forced to appear after the formation of the 
central singularity. In that case, the apparent horizon 
curve begins at the central singularity $r=0$ at $t=t_0$ 
and increases with increasing $r$, moving to the future.
Then we have $t_{ah} > t_0$ for $r > 0 $ near the center. 
The behaviour of outgoing families of null geodesics 
has been analyzed in detail in the case
when $\chi(0)>0$ and we can see that the geodesics 
terminate at the singularity in the past. Thus timelike 
and null geodesics come out from the singularity,
making it visible to external observers
(Joshi \& Dwivedi 1999).

One thus sees that it is the term $g_1$ in the 
stresses $p_\theta$ which decides either the black hole or 
naked singularity as the final fate for the collapse. We can choose 
it to be arbitrarily small, and it is then possible to 
see how introducing a generic tangential stress perturbation 
in the model would change drastically the final outcome 
of the collapse. For example, for all non-vanishing
tangential stresses with $g_0=0$ and $g_1<0$, even
the slightest perturbation of the Oppenheimer-Snyder-Datt
scenario, injecting a small tangential stress would result 
in a naked singularity. The space of all functions $g_1$ 
that make $\chi(0)$ positive, which includes all the 
strictly negative functions $g_1$, causes the collapse 
to end in a naked singularity. While this is an explicit 
example, it is by no means the only class. The important 
feature of this class is that it corresponds to a 
collapse model for a simple and straightforward perturbation 
of the Oppenheimer-Snyder-Datt spacetime metric.

In this case, the geometry near the centre can 
be written as,
\begin{equation}\label{pert}
ds^2=-(1-2g_1 r^3)dt^2+\frac{(v+rv')^2}{1+kr^2-2g_1 r^3}dr^2+r^2v^2d\Omega^2 \; ,
\end{equation}
The metric above satisfies the Einstein equations 
in the neighborhood of the center of the cloud when the 
function $g_1(v)$ is small and bounded. We can take 
$0<|g_1(v)|<\epsilon$, so that, the smaller we take the 
parameter $\epsilon$, the bigger will be the radius 
where the approximation is valid.
We can consider here the requirement that a realistic 
matter model should satisfy some energy conditions ensuring 
the positivity of mass and energy density.
The weak energy condition would imply restrictions 
on the density and pressure profiles. The energy density 
as given by the Einstein equation must 
be positive. Since $R$ is positive, to ensure positivity
of $\rho$ we require $F>0$ and $R'>0$. The choice of
positive $M(r)$, which clearly holds for $M_0>0$, 
and is physically reasonable, ensures positivity of the 
mass function. Here $R'>0$ is a sufficient condition 
for the avoidance of shell crossing singularities. 
The tangential stress can now be written, with 
$p_r=0$, and is given by 
\begin{equation}
p_\theta=\frac{1}{2}\frac{R}{R'}\rho\sigma'
\end{equation}
So the sign of the function $\sigma'$ would determine 
the sign of $p_\theta$. Positivity of $\rho+p_\theta$ is 
then ensured for small values of $r$ throughout collapse 
for any form of $p_\theta$. In fact, regardless of the values 
taken by $M$ and $g$, there will always be a neighbourhood 
of the center $r=0$ for which $|p_\theta|<\rho$ and 
therefore $\rho+p_\theta\geq0$.

What we see here is that, in the space of initial data
in terms of the initial matter densities and velocity 
profiles, any arbitrarily small neighborhood of the OSD 
collapse model, which is going to end as a black hole, 
contains collapse evolutions that go to naked singularities.
Such an existence of subspaces of collapse solutions, 
that go to a naked singularity rather than a black
hole, in the arbitrary vicinity of the OSD black hole, 
presents an intriguing situation. It gives an idea of the 
richness of structure present in the gravitation theory, and 
indicates the complex solution space of the Einstein equations
which are a complicated set of highly non-linear partial 
differential equations. What we see here is the existence of
classes of stress perturbations such that an arbitrarily 
small change from the OSD model is a solution going to a naked 
singularity.

This then provides an intriguing insight into 
the nature of cosmic censorship, namely that the collapse 
must necessarily be properly fine-tuned if it is to produce
a black hole only as the final endstate. Traditionally it 
was believed that the presence of stresses or pressures 
in the collapsing matter cloud would increase the chance of 
black hole formation, thereby ruling out dust models that 
were found to lead to a naked singularity as the collapse endstate.
It now becomes clear that this is actually not the case. 
The model described here not only provides a new class 
of collapses ending in naked singularities, but more importantly, 
shows how the bifurcation line that separates the phase space 
of `black hole formation' from that of `naked singularity 
formation' runs directly over the simplest and most studied 
of black hole scenarios such as the OSD model.

It has to be noted of course that the general issue
of stability and genericity of collapse outcomes has been 
a deep problem in gravitation theory, and requires 
mathematical breakthroughs as well as evolving further 
physical understanding of the collapse phenomenon. 
As noted above, this is again basically connected 
with the main difficulty of cosmic censorship itself, 
which is the issue of how to define censorship.
However, it is also clear from the 
discussion above, that consideration of various collapse 
models along the lines as discussed here does yield  
considerable insight and inputs in understanding 
gravitational collapse and its final outcomes.

\section{Spinning up a black hole and quantum aspects}

It is clear that the black hole and naked singularity 
outcomes of a complete gravitational collapse for a massive 
star are very different from each other physically, and 
would have quite different observational signatures. 
In the naked singularity case, if it occurs in nature, 
we have the possibility of observing the physical effects 
happening in the vicinity of the ultra-dense regions that form 
in the very final stages of collapse. However, in a black 
hole scenario, such regions are necessarily hidden
within the event horizon of gravity. The fact that a 
slightest stress perturbation of the OSD collapse could 
change the collapse final outcome drastically, as we 
noted in the previous section, changing it from black 
hole formation to a naked singularity, means that the 
naked singularity final state for a collapsing star 
must be studied very carefully to deduce its physical 
consequences, which are not well understood so far.

It is, however, widely believed that when we have
a reasonable and complete quantum theory of gravity
available, all spacetime singularities, whether naked 
or those hidden inside black holes, will be resolved away.   
As of now, it remains an open question if quantum 
gravity will remove naked singularities.
After all, the occurrence of spacetime singularities 
could be a purely classical phenomenon, and 
whether they are naked or covered should not be relevant,
because quantum gravity will possibly remove them 
all any way. But one may argue that looking at the 
problem this way is missing the real issue. 
It is possible that in a suitable quantum gravity theory 
the singularities will be smeared out, though this has 
not been realized so far. Also there are indications that 
in quantum gravity also the singularities may not 
after all go away.

In any case, the important and real issue is, 
whether the extreme strong gravity regions formed due 
to gravitational collapse are visible to faraway observers 
or not. It is quite clear that gravitational collapse 
would certainly proceed classically, at least till 
quantum gravity starts governing and dominating the 
dynamical evolution at scales of the order 
of the Planck length, {\it i.e.} till extreme gravity 
configurations have been already developed due to 
collapse. The key point is the visibility or 
otherwise of such ultra-dense regions 
whether they are classical or quantum (see Fig. 5).

\begin{figure}
\centerline{\includegraphics[width=9cm]{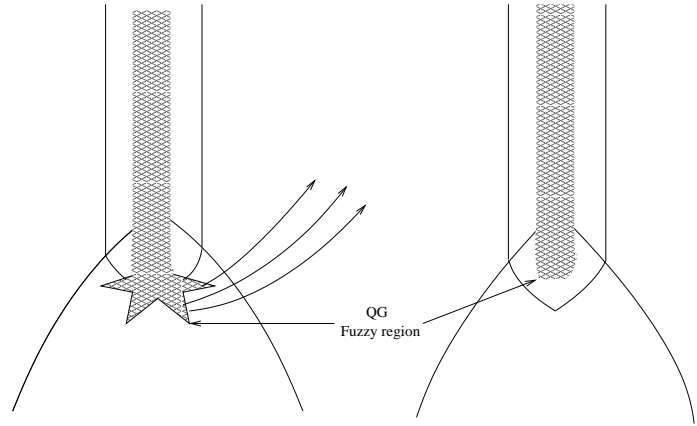}}
\caption{Even if the naked singularity is resolved by
the quantum gravity effects, the ultra-strong 
gravity region that developed in gravitational collapse 
will still be visible to external observers
in the Universe. \label{f:four}}
\end{figure}

What is important is, classical gravity implies 
necessarily the existence of ultra-strong gravity 
regions, where 
both classical and quantum gravity come into their own. 
In fact, if naked singularities do develop in gravitational 
collapse, then in a literal sense we come face-to-face 
with the laws of quantum gravity, whenever such an 
event occurs in the Universe (Wald 1997).

In this way, the gravitational collapse phenomenon 
has the potential to provide us with a possibility of 
actually testing the laws of quantum gravity.
In the case of a black hole developing in the 
collapse of a finite sized object such as a massive star, 
such strong gravity regions are necessarily hidden 
behind an event horizon of gravity, and this would happen
well before the physical conditions became extreme
near the spacetime singularity. 
In that case, the quantum effects, even if they caused 
qualitative changes closer to singularity, will be 
of no physical consequence. This is because no causal 
communications are then allowed from such regions. On 
the other hand, if the causal structure were that 
of a naked singularity, then communications from 
such a quantum gravity dominated extreme curvature 
ball would be visible in principle. This will be so
either through direct physical processes near a 
strong curvature naked singularity, or via the 
secondary effects, such as the shocks produced in 
the surrounding medium.

Independently of such possibilities connected with
gravitational collapse as above, let us suppose that 
the collapse terminated in a black hole. It is generally 
believed that such a black hole will be described by 
the Kerr metric. A black hole, 
however, by its very nature accretes matter from the 
surrounding medium or from a companion star. In that case, 
it is worth noting here that there has been an active 
debate in recent years about whether 
a black hole can survive as it is, when it accretes 
charge and angular momentum from the surrounding medium.

The point is, there is a constraint in this case 
for the horizon to remain undisturbed, namely that 
the black hole must not contain too much of charge  and
it should not spin too fast. Otherwise, 
the horizon cannot be sustained. It will breakdown 
and the singularity within will become visible.
The black hole may have formed with small enough 
charge and angular momentum to begin with; however, 
we have the key astrophysical process of accretion from 
its surroundings, of the 
debris and outer layers of the collapsing star. 
This matter around the black hole will fall 
into the same with great velocity, which could be classical 
or quantized, and with either charge or angular momentum
or perhaps both. Such in-falling particles could 
`charge-up' or `over-spin' the black hole, thus eliminating 
the event horizon. Thus, the very fundamental 
characteristic of a black hole, namely its trait of 
gobbling up the matter all around it and continuing to grow
becomes its own nemesis and a cause of 
its own destruction.

Thus, even if a massive star collapsed into a 
black hole rather than a naked singularity, important issues 
remain such as the stability against accretion of
particles with charge or large angular momentum, and whether 
that can convert the black hole into a naked singularity by 
eliminating its event horizon. Many researchers have claimed 
this is possible, and have given models to create naked 
singularities this way. But there are others who claim 
there are physical effects which would save the black hole 
from over-spinning this way and destroying itself, and the issue is 
very much open as yet. The point is, in general, the 
stability of the event horizon and the black hole 
continues to be an important issue for black 
holes that developed in gravitational collapse.
For a recent discussion on some of these issues, 
we refer to Matsas \& da Silva (2007), Matsas et al. (2009), 
Hubeny (1999), Hod (2008), Richartz \& Saa (2008), 
Jacobson \& Sotiriou (2009, 2010a,b), Barausse, Cardoso \& Khanna 
(2010), and references therein.

The primary concern of the cosmic censorship 
hypothesis is the formation of black holes 
as collapse endstates. Their stability, as discussed
above, is only a secondary issue. Therefore, what this 
means for cosmic censorship is that the collapsing massive star 
should not retain or carry too much charge or spin; 
otherwise it could necessarily end up as a naked 
singularity, rather than a black hole.

\section{Predictability, Cauchy horizons and all that}
A concern sometimes expressed is that if naked 
singularities occurred as the final fate of gravitational 
collapse, that would break the predictability in the 
spacetime. A naked singularity is characterized by the 
existence of light rays and particles that emerge from the 
same. Typically, in all the collapse models discussed 
above, there is a family of future directed non-spacelike 
curves that reach external observers, and when extended
in the past these meet the singularity. 
The first light ray that comes out from the singularity
marks the boundary of the region that can be predicted 
from a regular initial Cauchy surface in the spacetime,
and that is called the Cauchy horizon for the spacetime.
The causal structure of the spacetime would differ
significantly in the two cases, when there is a Cauchy
horizon and when there is none. A typical 
gravitational collapse to a naked singularity, with
the Cauchy horizon forming is shown in Fig. 6.

\begin{figure}
\centerline{\includegraphics[width=9cm]{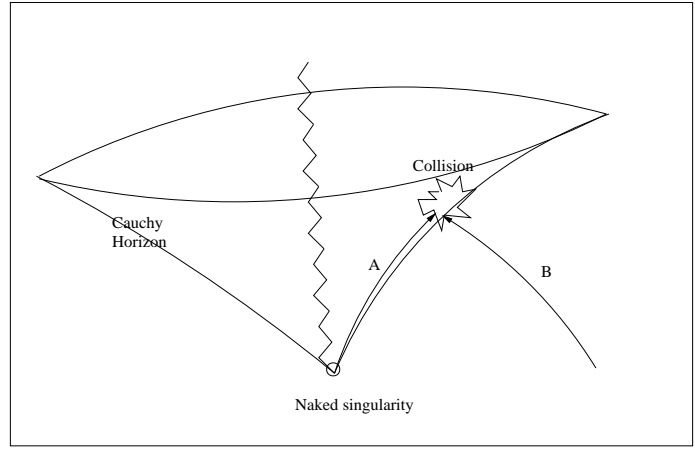}}
\caption{The existence of a naked singularity is
typically characterized by existence of a Cauchy horizon
in the spacetime. Very high energy particle collisions
can occur close to such a Cauchy horizon. \label{f:four}}
\end{figure}

The point here is, given a regular initial data on a 
spacelike hypersurface, one would like to predict the future
and past evolutions in the spacetime for all times
(see for example, Hawking \& Ellis (1973)
for a discussion).
Such a requirement is termed the global hyperbolicity 
of the spacetime. A globally hyperbolic spacetime is a fully
predictable universe. It admits a Cauchy surface, the data
on which can be evolved for all times in the past as well
as in future. Simple enough spacetimes such as the Minkowski
or Schwarzschild are globally hyperbolic, but the Reissner-Nordstrom
or Kerr geometries are not globally hyperbolic. For further 
details on these issues, we refer to 
Hawking \& Ellis (1973) or Joshi (2008).

Here we would like to mention certain recent intriguing 
results in connection to the existence of a Cauchy horizon 
in a spacetime when there is a naked singularity resulting as 
final fate of a collapse. Let us suppose the collapse resulted 
in a naked singularity. In that case, there are classes
of models where there will be an outflow of energy and 
radiations of high velocity particles close to the Cauchy 
horizon, which is a null hypersurface in the spacetime.
Such particles, when they collide with incoming particles,
would give rise to a very high center of mass energy 
of collisions.
The closer we are to the Cauchy horizon, higher is the 
center of mass energy of collisions. In the limit of
approach to the Cauchy horizon, these energies approach 
arbitrarily high values and could reach the 
Planck scale energies (see for example, Patil \& Joshi 2010, 2011a,b; 
Patil, Joshi \& Malafarina 2011).

It has been observed recently that in the vicinity
of the event horizon for an extreme Kerr black hole, 
if the test particles arrive with fine-tuned velocities, 
they could undergo very high energy collisions with other 
incoming particles. In that case, the possibility 
arises that one could see Planck scale physics
or ultra-high energy physics effects 
near the event horizon, given suitable circumstances
(Banados, Silk \& West 2009; Berti et al. 2009; 
Jacobson \& Sotiriou 2010a,b; Wei et al. 2010;
Grib \& Pavlov 2010; Zaslavskii, 2010).

What we mentioned above related to the particle
collisions near Cauchy horizon is similar to the situation
where the background geometry is that of a naked 
singularity. These results could mean that in strong 
gravity regimes, such as those of black holes or naked 
singularities developing in gravitational collapse, 
there may be a possibility to observe
ultra-high energy physics effects, which would 
be very difficult to see in the near future 
in terrestrial laboratories.   

While these phenomena give rise to the prospect of
observing Planck scale physics near the Cauchy horizon 
in the gravitational collapse
framework, they also raise the following intriguing question.
If extreme high energy collisions do take place very close
to the null surface which is the Cauchy horizon, then
in a certain sense it is essentially equivalent to creating 
a singularity at the Cauchy horizon. In that case, all or
at least some of the Cauchy horizon would be converted into
a spacetime singularity, and would effectively mark 
the end of the spacetime itself. In this case, the 
spacetime manifold terminates 
at the Cauchy horizon, whenever a naked singularity is 
created in gravitational collapse. Since the Cauchy horizon
marks in this case the boundary of the spacetime itself, 
predictability is then restored for the spacetime, 
because the rest of the spacetime below and in the 
past of the horizon 
was predictable before the Cauchy horizon formed.

\section{Future perspectives}

We have pointed out in the considerations here 
that the final fate of a
massive star continues to be a rather exciting research 
frontier in black hole physics and gravitation 
theory today. The outcomes here will be fundamentally
important for the basic theory and astrophysical 
applications of black hole physics, and for modern
gravitation physics. We highlighted certain key 
challenges in the field, and also several recent 
interesting developments were reviewed. Of course, 
the issues and the list given here are by no means complete 
or exhaustive in any manner, and there are several 
other interesting problems in the field as well.

In closing, as a summary, we would like to mention 
here a few points which we think require the most immediate 
attention, and which will have possibly maximum 
impact on future development in the field.

1. The genericity of the collapse outcomes, for 
black holes and naked singularities need to
be understood very carefully and in further detail. 
It is by and large well-accepted now, that the general 
theory of relativity does allow and gives rise to 
both black holes and naked singularities as the final 
outcome of continual gravitational collapse, evolving 
from a regular initial data, and under reasonable 
physical conditions.   
What is not fully clear yet is the distribution 
of these outcomes in the space of all allowed outcomes
of collapse. The collapse models discussed above 
and considerations we gave here would be of some 
help in this direction, and may throw some light on 
the distribution of black holes and naked singularity 
solutions in the initial data space.

2. Many of the models of gravitational collapse
analyzed so far are mainly of spherical symmetric collapse.
Therefore, the non-spherical collapse needs to be 
understood in a much better manner. While there are 
some models which illustrate what the departures
from spherical symmetry could do (see e.g.
Joshi \& Krolak 1996),
on the whole, not very much is known for non-spherical
collapse. Probably numerical relativity could be
of help in this direction
(see for example 
Baiotti \& Rezzolla 2006),
for a discussion on the evolving developments as
related to applications of numerical methods
to gravitational collapse issues).
Also, another alternative would
be to use global methods to deal with the spacetime 
geometry involved, as used in the case of singularity 
theorems in general relativity.

3. At the very least, the collapse models 
studied so far do help us gain much insight into the 
structure of the cosmic censorship, whatever 
final form it may have.
But on the other hand, there have also been attempts 
where researchers have explored the physical applications 
and implications of the naked singularities investigated 
so far (see e.g. 
Harada, Iguchi \& Nakao 2000, 2002; 
Harada et al. (2001) and also references therein).

If we could find astrophysical applications of 
the models that predict naked singularities, 
and possibly try to test the same through 
observational methods and the signatures predicted,
that could offer a very interesting avenue to get 
further insights into this problem as a whole. 

4. An attractive recent possibility in that 
regard is to explore the naked singularities as 
possible particle accelerators as we pointed 
out above.  
	
Also, the accretion discs around a naked 
singularity, wherein the matter particles are 
attracted towards or
repulsed away from the singularities with great
velocities could provide an excellent venue to test 
such effects and may lead to predictions of 
important observational signatures to distinguish
the black holes and naked singularities in
astrophysical phenomena
(see Kovacs \& Harko 2010; Pugliese, Quevedo \& Ruffini 2011).

5. Finally, further considerations of quantum 
gravity effects in the vicinity of naked singularities,
which are super-ultra-strong gravity regions,  
could yield intriguing theoretical insights into
the phenomenon of gravitational collapse 
(Goswami, Joshi \& Singh 2006).

\section*{Acknowledgments}
Over past years, discussions with many colleagues and 
friends have contributed greatly to shape my understanding 
of the questions discussed here. In particular, I would like to
thank I. H. Dwivedi, N. Dadhich, R. Goswami, T. Harada, 
S. Jhingan, R. Maartens, K. Nakao, R. Saraykar, T. P. Singh, 
R. Tavakol, and also many other friends with whom I 
have extensively discussed these issues.
Fig. 4 here is from Mena, Tavakol \& Joshi (2000),
and Fig. 6 is from Patil, Joshi \& Malafarina (2011). The rest of the figures
are from Joshi (2008).

\label{lastpage}
\end{document}